\documentclass[english,notitlepage,nofootinbib]{revtex4-1}
\usepackage[T1]{fontenc}
\usepackage[latin9]{inputenc}
\usepackage{color}
\usepackage{babel}
\usepackage{amsmath}
\usepackage{cancel}
\usepackage{stackrel}
\usepackage{graphicx}
\usepackage[unicode=true,pdfusetitle,
 bookmarks=true,bookmarksnumbered=false,bookmarksopen=false,
 breaklinks=false,pdfborder={0 0 1},backref=false,colorlinks=true]
 {hyperref}
\hypersetup{
 allcolors=blue}

\makeatletter
\pdfoutput=1

\usepackage{color}
\usepackage{babel}

\usepackage{babel}

\makeatother

\begin{document}

\title{KLT-Like Behaviour of Inflationary Graviton Correlators}

\author{Shing Yan Li\textsuperscript{1}, Yi Wang\textsuperscript{1,2},
Siyi Zhou\textsuperscript{1,2}}
\email{syliah@connect.ust.hk, phyw@ust.hk, szhouah@connect.ust.hk}

\affiliation{\textsuperscript{1}Department of Physics, The Hong Kong University
of Science and Technology, Clear Water Bay, Kowloon, Hong Kong, P.R.China}

\affiliation{\textsuperscript{2}Jockey Club Institute for Advanced Study, The
Hong Kong University of Science and Technology, Clear Water Bay, Kowloon,
Hong Kong, P.R.China}
\begin{abstract}
We use the spinor helicity formalism to study KLT-like relations for
the inflationary graviton four-point correlation function. New features
are observed in this correlation function compared to the graviton
scattering amplitude in flat \textcolor{black}{spacetime}. After obtaining
the general momentum dependence, collinear, squeezed and collapsed
limits are considered to further study the features of the correlation
function, and the relation to the corresponding flat space scattering
amplitude. 
\end{abstract}
\maketitle

\section{Introduction}

Inflation \citep{Guth:1980zm,Linde:1981mu,Albrecht:1982wi,Starobinsky:1982ee}
is a hypothetical high-energy process in the very early universe.
It is believed that quantum fluctuations of various fields during
inflation exit the horizon, being seeds of complicated structures
of galaxies nowadays. The fluctuations involve the interactions between
inflatons, which are hypothetical particles produced during inflation
responsible for the \textcolor{black}{primordial} density perturbations.
There are also interactions between inflatons and gravitons, and purely
between gravitons, which produced primordial gravitational waves.
Interactions between scalars, especially in the soft momenta limit,
are well studied. Many interesting properties are discovered and compared
to those in flat spacetime \citep{Assassi:2012zq,Chu:2018ovy}.

With the recent detection of the gravitational waves produced by the
\textcolor{black}{black hole} merger \citep{Abbott:2016blz} and neutron
star merger \citep{TheLIGOScientific:2017qsa}, as well as the forthcoming
LISA experiments about primordial gravitational waves, the gravitational
wave astronomy is attracting more and more attentions. It is hoped
that in the future we will also have results about primordial gravitational
waves \cite{Abazajian:2016yjj,Li:2017drr} in terms of correlation
functions with different types of polarizations, thus it is desiring
to study the properties of such correlation functions during inflation.
In this paper we focus on the interactions between gravitons only.

The fluctuations in inflation are characterised by correlation functions.
The two-point functions, or power spectrum, represent the Gaussian
perturbations without interactions. Higher-point functions represent
more special features of perturbations, known as non-Gaussianities,
when interactions are present \citep{Chen:2010xka}. The three-point
functions in the slow-roll model of single minimally coupled scalar
field is first calculated in \citep{Maldacena:2002vr}. The correlation
functions are similar to scattering amplitudes considered in particle
physics. \textcolor{black}{However, particle physicists usually consider
amplitudes in flat spacetime, while we are in spacetime during inflation,
or a nearly de Sitter spacetime, and for simplicity we only consider
a purely de Sitter spacetime. The symmetries present in these two
spacetimes are different, and it is interesting to study how the mathematical
structures of the correlation functions are affected when the symmetry
group changes.} It would be nice if there exist some non-trivial relationships
between inflation and flat spacetime, and the well-studied interesting
properties of amplitudes can be generalized to our context.

One of the wonderful properties of graviton scattering amplitudes
in flat spacetime can be schematically written as ``$\mathrm{Gravity}=\mathrm{Gauge}^{2}$''
\citep{Kawai:1985xq}. Here ``gauge'' means Yang-Mills theory, which
is the gauge theory of gluons or strong interactions. It means that
the amplitudes of graviton scattering are roughly the square of those
of gluon scattering, due to a simple idea that a massless spin 2 graviton
is double copy of massless spin 1 gluons. There is the BCJ conjecture
of duality between colour factors and kinematic factors in gluon amplitudes,
so that we can just replace colour factors in gluon amplitudes by
copies of kinematic factors to obtain graviton amplitudes \citep{Bern:2008qj,Tye:2010dd}.
In string theory this fact is precisely described by KLT relations
\citep{Kawai:1985xq}, which relates closed strings and open strings
as ``$\mathrm{Closed}=\mathrm{Open}^{2}$'', and we expect they
will have projections on flat spacetime quantum field theory. For
example, the KLT relation for four-point amplitudes is 
\begin{equation}
M\left(1234\right)=nA\left(1234\right)A\left(1243\right)~,\label{eq:1}
\end{equation}

\noindent where $M$ and $A$ denote for graviton and \textcolor{black}{colour-stripped}
gluon amplitudes respectively, and $n$ is some kinematic factor which
is not important in this paper. The numbers in the brackets label
the external particles, and the orders of numbers represent the arrangement
of the particles in clockwise direction.

One would expect similar relations can also be found in inflationary
context. \textcolor{black}{While we indeed find a relation similar
to KLT relations in our context, there are some extra terms in the
relation, which may correspond to features appearing in de Sitter
spacetime only. They make the KLT structure very opaque. Especially,
there are terms which do not have the form of $\mathrm{Gauge}^{2}$
clearly. Therefore, we would like to emphasize that the relations
in this paper are only preliminary steps towards a complete KLT-like
relation in inflation, and further studies are needed to solidify
the relations.}

If we use Feynman diagrams to calculate gluon and graviton amplitudes,
extremely complicated expressions with thousands of terms are obtained
but they can be grouped into a single term when the amplitude is maximally
helicity violating. The result is known as Parke-Taylor formula \citep{Parke:1986gb}.
This means there are some symmetries hidden in the ordinary expressions.
In the literature of scattering amplitudes, this feature is revealed
by extra tools such as spinor helicity formalism \citep{Berends:1981rb,DeCausmaecker:1981jtq,Kleiss:1985yh,Gastmans:1990xh,Xu:1986xb,Gunion:1985vca},
BCFW recursion relations \citep{Britto:2005fq} and so on (see nice
reviews \citep{Dixon:1996wi,Elvang:2013cua}). Here we will also study
spinor helicity formalism in details to see its generalization to
inflation. \textcolor{black}{In addition, the main result of this
paper, the KLT-like behaviour, is derived with this formalism, which
can greatly simplify the calculations needed.}

This paper is organised as follows. In Section \ref{sec:2} we review
general properties of the gravitons, and the computations of correlation
functions of graviton interactions. In Section \ref{sec:3} we discuss
the generalization of spinor helicity formalism to inflationary spacetime,
and use it to compute a four-point function. In this procedure we
derive a KLT-like relation. In Section \ref{sec:4} we describe attempts
to interpret such behaviour and point out its significance, by considering
various limits of momentum configurations. We conclude and talk about
possible extension of this work in Section \ref{sec:5}.

\section{Graviton Correlators in Inflation\label{sec:2}}

To begin with, we review the calculation of three-point functions
of three gravitons in inflation. The calculation has been done \citep{Maldacena:2002vr}
and we just recall some key points which are useful to the analysis
below. For simplicity, we consider the model of single minimally coupled
scalar field. We use the $\left(-,+,+,+\right)$ metric convention.
Then the action is given by 
\begin{equation}
S=\int d^{4}x\,\sqrt{-g}\left(\frac{R}{2}-\frac{1}{2}g^{\mu\nu}\partial_{\mu}\phi\partial_{\nu}\phi-V\left(\phi\right)\right)~,
\end{equation}

\noindent where $\phi$ is the inflaton field which can be decomposed
as background and perturbations as $\phi=\bar{\phi}+\delta\phi$.
One can expand the action to arbitrary order using ADM formalism \citep{Baumann:2009ds}
\begin{equation}
g_{00}=-N^{2}+g_{ij}N^{i}N^{j},\:g_{0i}=g_{ij}N^{j},\:g_{ij}=a^{2}\exp\left(h\right)_{ij}=e^{2Ht}\exp\left(h\right)_{ij}~,
\end{equation}

\noindent where $H$ is the Hubble parameter, $N$ and $N^{i}$ are
lapse and shift functions. To study graviton fluctuations only, in
the traceless and transverse gauge of the gravitons, we can set $N=1$
and $N^{i}=0$ for the third order action.

We quantize the graviton field by the second order action 
\begin{equation}
S_{2}=\frac{1}{8}\int d\tau d^{3}x\,a^{2}\left(h_{ij}'h_{ij}'-\partial_{l}h_{ij}\partial_{l}h_{ij}\right)~,
\end{equation}

\noindent where $h_{ij}$ is the graviton field from Ricci scalar
and we set $M_{Pl}=\left(8\pi G\right)^{-1/2}=1$. From now on we
only use conformal time. Now we decompose the field to scalars by
polarization tensors and quantize the scalars 
\[
h_{ij}\left(\mathbf{k}\right)=\underset{s=\pm}{\sum}\epsilon_{ij}^{s}\left(\mathbf{k}\right)h_{\mathbf{k}}^{s}~,
\]
\begin{equation}
h_{\mathbf{k}}^{s}=\frac{H}{\sqrt{2k^{3}}}\left(1+ik\tau\right)e^{-ik\tau}a_{\mathbf{k}}^{s}+\frac{H}{\sqrt{2k^{3}}}\left(1-ik\tau\right)e^{ik\tau}a_{\mathbf{-k}}^{s\dagger}~.\label{eq5}
\end{equation}

Here $a^{\dagger}$ and $a$ are the creation and annihilation operators.
We use circular polarization and choose the traceless and transverse
gauge. Thus, the polarization tensors satisfy 
\begin{equation}
\epsilon_{ij}^{s}=\epsilon_{ji}^{s},\:\partial_{i}\epsilon_{ij}^{s}=k_{i}\epsilon_{ij}^{s}=0,\:\epsilon_{ii}^{s}=0~.
\end{equation}

We also normalize the polarization tensors by $\epsilon_{ij}^{s}\epsilon_{ij}^{*s'}=4\delta_{ss'}$.
Now the three-point interaction is determined by the third order action,
which is given by \citep{Maldacena:2002vr,Maldacena:2011nz,Fu:2015vja}
\begin{equation}
S_{3}=\frac{1}{8}\int d\tau d^{3}x\,a^{2}\left(h_{kl}\partial_{k}h_{ij}-2h_{ik}\partial_{k}h_{jl}\right)\partial_{l}h_{ij}~.
\end{equation}

We can already expect some relations between this action and that
of flat spacetime since the integrand is just the one in flat spacetime
multiplied by $a^{2}$ \citep{Bern:1999ji}. Let us also discuss some
features of the correlator first. Due to momentum conservation, the
correlator must take the form 
\begin{equation}
\left\langle h_{\mathbf{k}_{1}}^{s_{1}}h_{\mathbf{k}_{2}}^{s_{2}}h_{\mathbf{k}_{3}}^{s_{3}}\right\rangle =\left(2\pi\right)^{3}\delta^{3}\left(\mathbf{k}_{1}+\mathbf{k}_{2}+\mathbf{k}_{3}\right)\left\langle h_{\mathbf{k}_{1}}^{s_{1}}h_{\mathbf{k}_{2}}^{s_{2}}h_{\mathbf{k}_{3}}^{s_{3}}\right\rangle '~.
\end{equation}

Here we define the symbol ``prime'' to be the correlation function
with the momentum conservation delta function removed. Since the energy
of the particles is no longer conserved, we just have 3-dimensional
delta function, which is different from that in flat spacetime. Next,
we note that both terms in the interaction have the form of $h\partial h\partial h$
and we can factorize the scalar fields out, remaining time-independent
products between tensors. Therefore if we only consider three-point
interactions, each contribution to the correlation functions from
each diagram can be written schematically as 
\begin{equation}
\left\langle h_{\mathbf{k}_{1}}^{s_{1}}...h_{\mathbf{k}_{n}}^{s_{n}}\right\rangle _{i}'=\left(\mathrm{scalar}\:\mathrm{part}\right)\left(\mathrm{tensor}\:\mathrm{part}\right)~,\label{eq:10}
\end{equation}

\noindent where the scalar part is unchanged when we replace the graviton
fields by inflaton-like fields $\delta\phi$ and consider a hypothetical
model with interaction $H_{I}'=-\frac{1}{8}\int d^{3}x\,a^{2}\delta\phi^{3}$.
This part can also be studied by some scattering amplitude technique
using the method developed in \citep{Arkani-Hamed:2017fdk}. Here
$i$ represents the diagram we are calculating.

\textcolor{black}{Now we demonstrate this fact by explicit calculations.
The inflationary correlation functions are calculated by the in-in
formalism \citep{Weinberg:2005vy} (see also \citep{Chen:2010xka,Wang:2013eqj,Chen:2017ryl})
\begin{equation}
\left\langle Q\right\rangle =\left\langle \left[\bar{T}e^{i\int_{-\infty}^{0}d\tau\,H_{I}\left(\tau\right)}\right]Q\left[Te^{-i\int_{-\infty}^{0}d\tau\,H_{I}\left(\tau\right)}\right]\right\rangle ~,\label{eq11}
\end{equation}
}

\noindent \textcolor{black}{where $H_{I}$ is the interaction Hamiltonian,
$\bar{T}$ and $T$ are anti-time-ordering and time-ordering operators
respectively. We then write down
\begin{equation}
H_{I}=-\frac{1}{8}\int d^{3}x\,a^{2}\left(h_{kl}\partial_{k}h_{ij}-2h_{ik}\partial_{k}h_{jl}\right)\partial_{l}h_{ij}~.
\end{equation}
}

\textcolor{black}{Now substitute $H_{I}$ and Equation (\ref{eq5})
into Equation (\ref{eq11}). We consider one term $X$ in the expansion
of the exponentials in one diagram. Since we only have interactions
in the form of $h\partial h\partial h$, in momentum space $X$ can
be decomposed schematically into
\begin{equation}
X=\sum\left(\left(\prod\int d\tau\prod-\frac{1}{8}\int d^{3}x\,a^{2}\prod k^{i}\right)\left\langle h_{\mathbf{k}_{1}}^{s_{1}}...h_{\mathbf{k}_{n}}^{s_{n}}h_{ij}...\right\rangle _{c}\right).
\end{equation}
}

\textcolor{black}{The expectation values $\left\langle h_{\mathbf{k}_{1}}^{s_{1}}...h_{\mathbf{k}_{n}}^{s_{n}}h_{ij}...\right\rangle _{c}$
consist of $h_{\mathbf{k}}^{s}$ and $h_{ij}$ only, and the numbers
of them are the same in each term in the sum. The subscript $c$ means
that we only consider the contribution from one certain way of contractions,
depending on the diagram we calculate. Note that only operators with
the same polarization have non-zero contractions, and the contraction
$\widehat{h_{\mathbf{k}}^{s}h_{\mathbf{k'}}^{s}}$ is independent
of $s$. Therefore it can be replaced by contraction between two scalar
fields $\widehat{\delta\phi_{\mathbf{k}}\delta\phi_{\mathbf{k'}}}$.
We also have
\[
\widehat{h_{\mathbf{k}}^{s}h_{ij}\left(\mathbf{k'}\right)}=\epsilon_{ij}^{s}\left(\mathbf{k'}\right)\widehat{h_{\mathbf{k}}^{s}h_{\mathbf{k'}}^{s}}=\epsilon_{ij}^{s}\left(\mathbf{k'}\right)\widehat{\delta\phi_{\mathbf{k}}\delta\phi_{\mathbf{k'}}}\;,
\]
}

\textcolor{black}{
\begin{equation}
\widehat{h_{ij}\left(\mathbf{k}\right)h_{lm}\left(\mathbf{k'}\right)}=\underset{s=\pm}{\sum}\epsilon_{ij}^{s}\left(\mathbf{k}\right)\epsilon_{lm}^{s}\left(\mathbf{k}\right)\widehat{h_{\mathbf{k}}^{s}h_{\mathbf{k'}}^{s}}=\widehat{\delta\phi_{\mathbf{k}}\delta\phi_{\mathbf{k'}}}\underset{s=\pm}{\sum}\epsilon_{ij}^{s}\left(\mathbf{k}\right)\epsilon_{lm}^{s}\left(\mathbf{k}\right)\;.
\end{equation}
}

\textcolor{black}{Therefore all contractions, and thus $\left\langle h_{\mathbf{k}_{1}}^{s_{1}}...h_{\mathbf{k}_{n}}^{s_{n}}h_{ij}...\right\rangle _{c}$
can be factorized into scalar parts and tensor parts. Moreover, all
such expectation values have the same scalar part $\prod\widehat{\delta\phi\delta\phi}$,
since we have specified the diagram we calculate. Now consider back
the whole sum. In each term in the summation, we identity $\prod\int d\tau\prod-\frac{1}{8}\int d^{3}x\,a^{2}\prod\widehat{\delta\phi\delta\phi}$
to be the scalar parts, and $\prod k^{i}\sum\left(\prod\epsilon_{ij}^{s}\right)$
to be the tensor parts. The tensor parts can be factorised out from
the integrations. On the other hand, it is clear that the scalar parts
are the same for all terms. We then reach
\begin{equation}
X=\left(\prod\int d\tau\prod-\frac{1}{8}\int d^{3}x\,a^{2}\prod\widehat{\delta\phi\delta\phi}\right)\sum\left(\prod k^{i}\prod\epsilon_{ij}^{s}\right).
\end{equation}
}

\textcolor{black}{Now consider the contribution from a diagram $\left\langle h_{\mathbf{k}_{1}}^{s_{1}}...h_{\mathbf{k}_{n}}^{s_{n}}\right\rangle _{i}'$.
It is the sum of such $X$ in which only the integral operators $\prod\int d\tau$
vary. Therefore in this sum the tensor part stays the same and we
conclude}

\textcolor{black}{
\begin{equation}
\left\langle h_{\mathbf{k}_{1}}^{s_{1}}...h_{\mathbf{k}_{n}}^{s_{n}}\right\rangle _{i}'=\sum\left(\prod\int d\tau\prod-\frac{1}{8}\int d^{3}x\,a^{2}\prod\widehat{\delta\phi\delta\phi}\right)\sum\left(\prod k^{i}\prod\epsilon_{ij}^{s}\right).
\end{equation}
}

\textcolor{black}{This justifies our claim. On the other hand, the
scalar part is equivalent to the expectation value $\left\langle \delta\phi\delta\phi...\delta\phi\right\rangle _{i}'$
with an effective Hamiltonian $H_{I}'$. From the above, $H_{I}'$
should consist of one integral operator $-\frac{1}{8}\int d^{3}x\,a^{2}$.
To get the correct number of $\widehat{\delta\phi\delta\phi}$, it
should be also a three-point interaction. Moreover, there is no derivatives
of fields. Therefore $H_{I}'=-\frac{1}{8}\int d^{3}x\,a^{2}\delta\phi^{3}$.}

\textcolor{black}{Be reminded that the tensor part still transforms
as a scalar, while it involves contracted products of tensors. We
observe that in this model this factorization only works for three-point
interactions. The tensor part is purely the kinematics of momenta
and polarizations, and is present in both de Sitter correlations and
flat spacetime amplitudes. On the other hand, the scalar part is really
the dynamics in de Sitter spacetime, and is not related to properties
of tensors. Note that one can make the same factorization for flat
spacetime amplitudes with three-point interactions. Therefore to study
the mathematical structures caused by tensor properties, for simplicity
we focus on comparing the tensor parts. Note that the pole structures
of correlation functions are always in the scalar part.}

Let us make an example of the above factorization. Applying the in-in
formalism to compute the three-point function, we get 
\begin{align}
\left\langle h_{\mathbf{k}_{1}}^{s_{1}}h_{\mathbf{k}_{2}}^{s_{2}}h_{\mathbf{k}_{3}}^{s_{3}}\right\rangle '= & -\frac{H^{4}}{16\left(k_{1}k_{2}k_{3}\right)^{3}}I\nonumber \\
 & \left(\left(\epsilon_{ii'}^{1}\epsilon_{jj'}^{2}\epsilon_{jj'}^{3}k_{i}^{2}k_{i'}^{2}-2k_{i}^{2}k_{j'}^{1}\epsilon_{ii'}^{1}\epsilon_{jj'}^{2}\epsilon_{ji'}^{3}\right)+2\:\mathrm{cyclic}\right)~,
\end{align}
\begin{equation}
I=-\left(k_{1}+k_{2}+k_{3}\right)+\frac{k_{1}k_{2}+k_{2}k_{3}+k_{3}k_{1}}{k_{1}+k_{2}+k_{3}}+\frac{k_{1}k_{2}k_{3}}{\left(k_{1}+k_{2}+k_{3}\right)^{2}}~.
\end{equation}

The second line is the tensor part. Below we will see that these six
terms can be grouped into a simple expression when the helicities
of the three gravitons are known.

One can calculate the four-point functions from graviton exchange
similarly, which has been done in \cite{Fu:2015vja}. We would see
the scalar part keeps the same for different scenarios \citep{Fu:2015vja,Seery:2008ax}.
An interesting point is that when we consider contributions from different
in-in contours and permutations separately, there are IR divergences.
However, they all cancel when we sum up the contributions.

For the tensor part, we just multiply the tensor parts of two three-point
vertices together, and sum up the helicity of internal graviton propagator.
\textcolor{black}{It is the same as that obtained by redoing the in-in
formalism.} We will treat this after introducing the spinor helicity
formalism. Once we have simplified the three-point vertex, we can
apply the results to calculate higher-point functions.

\textcolor{black}{Before starting the analysis, let us make a remark.
Below we only consider diagrams formed by three-point vertices only.
We are not talking higher-point vertices because they cannot be easily
transformed between 4 dimensions and 3 dimensions, that is, time derivatives
are involved in the Feynman rules \citep{Fu:2015vja,Bern:1999ji}.
Moreover, we cannot apply the factorization between scalar parts and
tensor parts to those vertices. On the other hand, contributions from
different channels contain different scalar parts, which are not our
main concern in this paper but cause great difficulty adding different
channels. Therefore as a preliminary step, we demonstrate the behaviour
of the four-point functions by considering only one channel with three-point
vertices only. From the results below, whether adding up different
channels can cause further simplification is quite non-trivial, and
we would like to leave it in further studies.}

\section{Spinor Helicity Formalism for Inflation\label{sec:3}}

Here we discuss the spinor helicity formalism for inflation. The spinor
helicity formalism for inflation is introduced in \citep{Maldacena:2011nz}
to simplify the above three-point functions. A review on that in flat
spacetime and the details of notations used here can be found in Appendix
\ref{sec:A}.

During inflation, we have nearly de Sitter background. For simplicity,
below we consider pure de Sitter background 
\begin{equation}
ds^{2}=\frac{1}{H^{2}\tau^{2}}\left(-d\tau^{2}+dx^{2}+dy^{2}+dz^{2}\right)~.
\end{equation}

Due to the expansion of the universe, the time translational symmetry,
which is present in flat spacetime, is broken and energy is not conserved
in general. However, we still have the 3-momentum conservation. Therefore
in inflation we usually work in 3-dimensional formalism i.e. considering
only spatial components of vectors and tensors. In this way we lose
the information about energy. In contrast, in flat spacetime, especially
in the spinor helicity formalism, we work in 4-dimensional formalism.
It means that some changes are needed to generalize the formalism
to 3 dimensions. Some formulas are also modified due to energy non-conservation.
As a result, some nice features in the spinor helicity formalism can
no longer be used.

\subsection{Modifications of the Formalism}

Although most results here were already obtained in \citep{Maldacena:2011nz},
here we work out more details of the formalism and emphasize some
points that were not mentioned. To begin with, the most simple generalization
is done by replacing 3-dimensional indices to 4-dimensional indices.
For example 
\begin{equation}
\epsilon_{ij}^{s}k_{i}k_{j}\rightarrow\epsilon_{\mu\nu}^{s}k^{\mu}k^{\nu}~,
\end{equation}

\noindent and the momentum vectors should be lightlike. We then define
\begin{equation}
k^{\mu}=\left(k,\boldsymbol{k}\right)~.
\end{equation}

Next, to force 4-dimensional results to be the same as 3-dimensional
results, we make sure the terms become zero when there are indices
being zero. Since we are considering products purely between polarization
tensors, and among polarization tensors and momenta, we require 
\begin{equation}
\epsilon_{0\nu}=\epsilon_{\mu0}=0~.
\end{equation}

To implement this, first we notice that under the gauge in Section
\ref{sec:2}, a polarization tensor can be written as direct product
of two vectors. We can set 
\begin{equation}
\epsilon_{+}^{\mu\nu}\left(p\right)=\frac{\left\langle p\right|\gamma^{\mu}\left|q\right]\left\langle p\right|\gamma^{\nu}\left|q\right]}{\left[qp\right]^{2}},\:\epsilon_{-}^{\mu\nu}\left(p\right)=\frac{\left\langle q\right|\gamma^{\mu}\left|p\right]\left\langle q\right|\gamma^{\nu}\left|p\right]}{\left\langle qp\right\rangle ^{2}}~,\label{eq16}
\end{equation}

\noindent where $q\neq p$ is the reference spinor. One can check
that Equation (\ref{eq16}) satisfies the remaining gauge and normalization
conditions. Here we can already see that gravitons are double copy
of gluons.

The zeroth components of the tensors are not zero in general. To ensure
they are always zero, we must choose $q$ to be 
\begin{equation}
\left|q\right]=\left|p\right\rangle ,\left|q\right\rangle =\left|p\right]~.
\end{equation}

This is not a convenient gauge to choose in flat spacetime, as it
breaks the Lorentz symmetry. Nevertheless, this choice allows us to
rewrite the graviton correlations into the spinor helicity formalism.

Now we cannot choose $q$ freely to simplify our calculations. Therefore,
many simplifications in flat spacetime no longer work. Below we will
also see that we have very different conclusions on correlators from
those in flat spacetime. We also introduce crossing between angle
brackets and square brackets, which makes our calculations even more
complicated. To be precise, we formally define the crossing as\footnote{Note that the crossing products defined here are different from those
in \citep{Elvang:2013cua}, which vanish by definition.} 
\begin{equation}
\left\langle pq\right]=\left\langle p\right|\gamma^{0}\left|q\right],\:\left[pq\right\rangle =\left[p\right|\gamma^{0}\left|q\right\rangle ~.
\end{equation}

One can then derive the following formulas: 
\begin{equation}
\left[pp\right\rangle =-\left\langle pp\right]=2p~,
\end{equation}
\begin{equation}
\left[pq\right\rangle \left[qp\right\rangle =2\left(pq+\mathbf{p}\cdot\mathbf{q}\right)~.\label{eq:20}
\end{equation}

Since we only have 3-momentum conservation now, the trick of momentum
conservation (see Appendix \ref{sec:A}) must be modified and there
are also variations of the trick due to the crossing products. For
example, we consider 3 momenta $\mathbf{k}_{1},\mathbf{k}_{2},\mathbf{k}_{3}$
with $\mathbf{k}_{1}+\mathbf{k}_{2}+\mathbf{k}_{3}=0$. We have 
\begin{equation}
\left\langle 12\right\rangle \left[23\right]=\left(k_{1}+k_{2}+k_{3}\right)\left\langle 13\right]~,\label{eq:21}
\end{equation}
\begin{equation}
\left\langle 12\right\rangle \left[23\right\rangle =\left(k_{1}+k_{2}-k_{3}\right)\left\langle 13\right\rangle ~,
\end{equation}
\begin{equation}
\left[12\right\rangle \left[23\right\rangle =\left(-k_{1}+k_{2}-k_{3}\right)\left[13\right\rangle ~.
\end{equation}

However, in flat spacetime, for instance, $\left\langle 12\right\rangle \left[23\right]=0$.
Therefore energy non-conservation makes our results more complicated.
Note that Equation (\ref{eq:21}) vanishes if the 3 momenta are on-shell
and energy is conserved i.e. $k_{1}+k_{2}+k_{3}=0$. Surely energy
of a particle cannot be negative, but since we are setting all external
particles to be incoming, we can analytically continue $k$ to $-k$
for outgoing particles.

The situation becomes even worse when there are more than 3 momenta.
For example, for the case of 4 momenta we have 
\begin{equation}
\left\langle 12\right\rangle \left[23\right\rangle =\left(k_{1}+k_{2}-k_{3}+k_{4}\right)\left\langle 13\right\rangle -\left\langle 14\right\rangle \left[43\right\rangle ~.
\end{equation}

As a result, when we consider correlation functions higher than three-point,
it is hard to eliminate the crossing products. It leads to some new
terms of the correlation functions which only appears in de Sitter
spacetime.

\subsection{Computation of Correlation Functions Using Spinor Helicity Formalism}

Here we compute the four-point functions using the formalism described
above. For the computation of three-point functions, which was briefly
done in \citep{Maldacena:2011nz}, see Appendix \ref{sec:B}. One
main message is that when we flip one $+$ helicity to $-$, we just
transform the original result by $\left.\right\rangle \rightarrow\left.\right]$
and $k\rightarrow-k$ and vice versa for corresponding graviton. Since
we only consider three-point vertices, this is true in general.

From now on when mentioning a correlation function, we refer to the
tensor part of it (see Equation (\ref{eq:10})), labeled as $\left\langle \mathrm{helicity}\right\rangle _{\mathrm{channel}}$,
unless otherwise specified. From the in-in formalism, the tensor part
of a higher-point function is just product of tensor parts of three-point
functions. Since the momentum 4-vectors are just defined artificially,
they can always be lightlike. Therefore, the three-point functions
in the product are just associated with momenta of the particles.
For simplicity, here we only consider four-point functions.

\begin{figure}
\begin{centering}
\includegraphics[width=0.3\paperwidth]{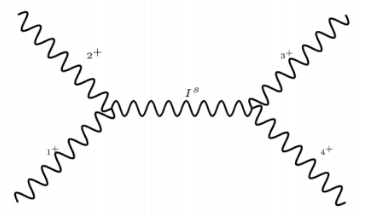} 
\par\end{centering}
\caption{$\left\langle 1^{+}2^{+}3^{+}4^{+}\right\rangle _{s}$}

\label{fig:1} 
\end{figure}

Consider the correlator $\left\langle 1^{+}2^{+}3^{+}4^{+}\right\rangle $,
where external legs of gravitons $1$ to $4$ are arranged clockwisely
in the diagram. Let the internal graviton have momentum $k_{I}^{i}$.
Note that the internal graviton can have $+$ or $-$ polarizations.
Here we only calculate the contribution from $s$ channel, see Figure
\ref{fig:1}. The contribution from other channels can be obtained
by simply permuting external gravitons. In addition, the results for
other combinations of helicities can be obtained by transformations
for flipping helicities.

The helicity index $s$ can be $+$ or $-$. We choose the internal
momentum incoming to the $12$ vertex and outgoing from the $34$
vertex. We call this choosing the direction of $k_{I}$. One can choose
the other direction and the result should be independent of the direction
we choose. That means the result should be even in $k_{I}$. To decompose
the diagram to two three-point functions, we analytically continue
the outgoing graviton in $34$ vertex to be incoming with opposite
momentum 4-vector and helicity \citep{Elvang:2013cua,Schwartz:2013pla}.
Therefore 
\begin{align}
\left\langle 1^{+}2^{+}3^{+}4^{+}\right\rangle _{s} & =\frac{1}{256\left(k_{1}k_{2}k_{3}k_{4}k_{I}^{2}\right)^{2}}[\left(\left\langle 12\right\rangle \left\langle 2I\right\rangle \left\langle I1\right\rangle \left\langle 34\right\rangle \left\langle 4I\right]\left[I3\right\rangle \left(k_{1}+k_{2}+k_{I}\right)\left(k_{3}+k_{4}+k_{I}\right)\right)^{2}\nonumber \\
 & +\left(\left\langle 34\right\rangle \left\langle 4I\right\rangle \left\langle I3\right\rangle \left\langle 12\right\rangle \left\langle 2I\right]\left[I1\right\rangle \left(k_{1}+k_{2}-k_{I}\right)\left(k_{3}+k_{4}-k_{I}\right)\right)^{2}]~.\label{eq:24}
\end{align}

Here we sum over helicity $s$. Using $\mathbf{k}_{I}=-\mathbf{k}_{1}-\mathbf{k}_{2}=\mathbf{k}_{3}+\mathbf{k}_{4}$
and applying Schouten's identity and momentum conservation repeatedly,
we get 
\begin{align}
\left\langle 2I\right\rangle \left\langle I1\right\rangle \left\langle 4I\right]\left[I3\right\rangle  & =\left(k_{3}-k_{4}+k_{I}\right)\left(k_{1}-k_{2}+k_{I}\right)\left\langle 23\right\rangle \left\langle 14\right\rangle \nonumber \\
 & +\left(-k_{3}+k_{4}+k_{I}\right)\left(k_{1}-k_{2}+k_{I}\right)\left\langle 24\right\rangle \left\langle 13\right\rangle +2k_{I}\left\langle 23\right\rangle \left\langle 24\right\rangle \left[21\right\rangle ~,
\end{align}
\begin{align}
\left\langle 4I\right\rangle \left\langle I3\right\rangle \left\langle 2I\right]\left[I1\right\rangle  & =\left(k_{3}-k_{4}-k_{I}\right)\left(k_{1}-k_{2}-k_{I}\right)\left\langle 23\right\rangle \left\langle 14\right\rangle \nonumber \\
 & +\left(-k_{3}+k_{4}-k_{I}\right)\left(k_{1}-k_{2}-k_{I}\right)\left\langle 24\right\rangle \left\langle 13\right\rangle -2k_{I}\left\langle 23\right\rangle \left\langle 24\right\rangle \left[21\right\rangle ~.
\end{align}

We can see that the transformation for flipping helicities also works
well for internal gravitons. The expressions seem to be much more
complicated. The purpose of this step is to express the four-point
functions in terms of external momenta and $k_{I}$ only. Nevertheless
there are angle-square bracket products remained, while we suppose
there are only angle bracket products when all 4 external gravitons
have $+$ helicity. In general, we would expect gravitons with $+$
helicity to associate with angle brackets, and gravitons with $-$
helicity to associate with square brackets. If there are terms which
do not follow this pattern, we call such terms anomalous terms. Note
that we just randomly choose the anomalous term $\left[21\right\rangle $
to appear, and one can derive similar expressions for all other possibilities.

Now we expand the square and use 
\begin{equation}
\left\langle 23\right\rangle \left\langle 24\right\rangle \left[21\right\rangle =\left(-k_{1}+k_{2}-k_{3}+k_{4}\right)\left\langle 23\right\rangle \left\langle 14\right\rangle +\left(-k_{1}+k_{2}+k_{3}-k_{4}\right)\left\langle 24\right\rangle \left\langle 13\right\rangle +\left\langle 13\right\rangle \left\langle 14\right\rangle \left[12\right\rangle ~,\label{eq:27}
\end{equation}

\noindent and Equation (\ref{eq:20}) to eliminate squares of angle-square
bracket products, we finally arrive 
\begin{align}
\left\langle 1^{+}2^{+}3^{+}4^{+}\right\rangle _{s} & =\frac{1}{256\left(k_{1}k_{2}k_{3}k_{4}k_{I}^{2}\right)^{2}}[2K_{c}\left(\left\langle 12\right\rangle \left\langle 23\right\rangle \left\langle 34\right\rangle \left\langle 41\right\rangle \right)\left(\left\langle 12\right\rangle \left\langle 24\right\rangle \left\langle 43\right\rangle \left\langle 31\right\rangle \right)\nonumber \\
 & +K_{s_{+}}\left(\left\langle 12\right\rangle \left\langle 23\right\rangle \left\langle 34\right\rangle \left\langle 41\right\rangle \right)^{2}+K_{s_{-}}\left(\left\langle 12\right\rangle \left\langle 24\right\rangle \left\langle 43\right\rangle \left\langle 31\right\rangle \right)^{2}\nonumber \\
 & -16\left(K_{a_{+}}\left\langle 12\right\rangle \left\langle 23\right\rangle \left\langle 34\right\rangle \left\langle 41\right\rangle +K_{a_{-}}\left\langle 12\right\rangle \left\langle 24\right\rangle \left\langle 43\right\rangle \left\langle 31\right\rangle \right)\left\langle 12\right\rangle \left\langle 23\right\rangle \left\langle 24\right\rangle \left\langle 34\right\rangle \left[21\right\rangle ]~,\label{eq:28}
\end{align}

\noindent where

\noindent 
\begin{align}
K_{c}= & \left(\left(\left(k_{1}+k_{2}+k_{I}\right)\left(k_{3}+k_{4}+k_{I}\right)\left(k_{1}-k_{2}+k_{I}\right)\right)^{2}+\left(\left(k_{1}+k_{2}-k_{I}\right)\left(k_{3}+k_{4}-k_{I}\right)\left(k_{1}-k_{2}-k_{I}\right)\right)^{2}\right)\nonumber \\
 & \left(k_{3}-k_{4}+k_{I}\right)\left(k_{3}-k_{4}-k_{I}\right)\nonumber \\
+ & 2k_{I}^{2}\left(\left(\left(k_{1}+k_{2}+k_{I}\right)\left(k_{3}+k_{4}+k_{I}\right)\right)^{2}+\left(\left(k_{1}+k_{2}-k_{I}\right)\left(k_{3}+k_{4}-k_{I}\right)\right)^{2}\right)\left(k_{1}-k_{2}+k_{I}\right)\left(k_{1}-k_{2}-k_{I}\right)~,
\end{align}
\begin{align}
K_{s_{\pm}} & =\left(\left(k_{1}+k_{2}+k_{I}\right)\left(k_{3}+k_{4}+k_{I}\right)\left(k_{1}-k_{2}+k_{I}\right)\left(\pm k_{3}\mp k_{4}+k_{I}\right)\right)^{2}\nonumber \\
 & +\left(\left(k_{1}+k_{2}-k_{I}\right)\left(k_{3}+k_{4}-k_{I}\right)\left(k_{1}-k_{2}-k_{I}\right)\left(\pm k_{3}\mp k_{4}-k_{I}\right)\right)^{2}~,
\end{align}
\begin{equation}
K_{a_{\pm}}=\left(k_{1}+k_{2}+k_{3}+k_{4}\right)k_{I}^{2}\left(\left(k_{1}-k_{2}\right)\left(k_{3}-k_{4}\right)\pm k_{I}^{2}\right)\left(\left(k_{1}+k_{2}\right)\left(k_{3}+k_{4}\right)+k_{I}^{2}\right)~.
\end{equation}

The expression is surely much longer than before, but the different
hidden roles of the products of tensors are now very clear. See discussions
in next section. Note that although $A\left(++++\right)=0$ for gluon
scattering, if we insert a higher-dimensional interaction, such as
$\mathcal{L}_{I}\sim\mathrm{tr}\,F_{\mu}^{\nu}F_{\nu}^{\lambda}F_{\lambda}^{\mu}$,
into the Yang-Mills theory, we would have $A\left(1^{+}2^{+}3^{+}4^{+}\right)\sim\left\langle 12\right\rangle \left\langle 23\right\rangle \left\langle 34\right\rangle \left\langle 41\right\rangle $
and $A\left(1^{+}2^{+}4^{+}3^{+}\right)\sim\left\langle 12\right\rangle \left\langle 24\right\rangle \left\langle 43\right\rangle \left\langle 31\right\rangle $.
In addition, the KLT relation for flat spacetime (see Equation (\ref{eq:1}))
is still true if we also insert the corresponding interaction $\mathcal{L}_{I}\sim R^{3}$
into Einstein gravity \citep{Cohen:2010mi}. Therefore the first line
in Equation (\ref{eq:28}) is analog to the KLT relation, while second
and third lines are some extra terms representing new features of
amplitudes that are not present in flat spacetime. For convenience,
we call the first, second and third lines to be cross terms, square
terms and anomalous terms respectively. \textcolor{black}{As a remark,
the existence of ``$\mathrm{Gravity}=\mathrm{Gauge}^{2}$'' is trivial
in our case since the polarization tensors are already square of polarization
vectors in gauge theory at the beginning of computation. However,
our main concern is how the ``squares'' look like in the KLT-like
relation, which remains non-trivial and interesting.}

Equation (\ref{eq:28}) also does not look symmetric. It is because
we have chosen $\left[21\right\rangle $ to be the anomalous term.
In principle, we can have a more symmetric form, but it turns out
that the biased form is more convenient for further calculations.
We have the freedom to choose the anomalous term to facilitate the
calculations.

Before analysing the results, let us make a remark. The angle-square
bracket products can be, in fact, expressed in terms of ordinary spinor
products. Using Equation (\ref{eq:27}) again, we have, for example
\[
\left\langle 23\right\rangle \left\langle 24\right\rangle \left[21\right\rangle =\frac{1}{2}\left(C\pm\sqrt{C^{2}-4\left(k_{1}-k_{2}+k_{I}\right)\left(k_{1}-k_{2}-k_{I}\right)\left\langle 23\right\rangle \left\langle 14\right\rangle \left\langle 24\right\rangle \left\langle 13\right\rangle }\right)~,
\]

\begin{equation}
C=\left(-k_{1}+k_{2}-k_{3}+k_{4}\right)\left\langle 23\right\rangle \left\langle 14\right\rangle +\left(-k_{1}+k_{2}+k_{3}-k_{4}\right)\left\langle 24\right\rangle \left\langle 13\right\rangle ~,\label{eq:35}
\end{equation}

\noindent and one can use other conditions to determine the sign above.
However, it remains hard to interpret the square root and the expression
becomes even longer. Therefore we prefer to keep the angle-square
bracket products.

\subsection{\textcolor{black}{Comparing with Yang-Mills Theory in Inflation}}

\textcolor{black}{So far we compare the graviton correlation functions
in inflation with the colour-stripped Yang-Mills scattering amplitudes
in flat spacetime. Although it suffices to show new features of the
KLT-like relation, to be precise we should also do the comparison
with the colour-stripped Yang-Mills correlation functions in inflation.
Note that in our simplest inflation model, we do not have Yang-Mills
interactions. However, since the behaviour of ``$\mathrm{Gravity}=\mathrm{Gauge}^{2}$''
is still clear for three-point functions (see Appendix \ref{sec:B}),
it is natural to write down
\begin{equation}
\left\langle 1^{+}2^{+}3^{+}\right\rangle ^{YM}=\frac{\left\langle 12\right\rangle \left\langle 23\right\rangle \left\langle 31\right\rangle }{4k_{1}k_{2}k_{3}}\left(k_{1}+k_{2}+k_{3}\right)\;,
\end{equation}
}

\noindent \textcolor{black}{and similar expressions for other combinations
of helicities. We then just repeat the calculations in the previous
section to obtain the corresponding four-point functions. To see whether
the analog of KLT relations when both sides are in inflation is non-trivial,
we just compare $\left\langle 1^{+}2^{+}3^{+}4^{+}\right\rangle _{s}$
with $\left\langle 1^{+}2^{+}3^{+}4^{+}\right\rangle _{s}^{YM}\left\langle 1^{+}2^{+}4^{+}3^{+}\right\rangle _{s}^{YM}$.
We then have
\begin{alignat}{1}
\left\langle 1^{+}2^{+}3^{+}4^{+}\right\rangle _{s}^{YM}\left\langle 1^{+}2^{+}4^{+}3^{+}\right\rangle _{s}^{YM} & =\frac{1}{256\left(k_{1}k_{2}k_{3}k_{4}k_{I}^{2}\right)^{2}}[2k_{c}\left(\left\langle 12\right\rangle \left\langle 23\right\rangle \left\langle 34\right\rangle \left\langle 41\right\rangle \right)\left(\left\langle 12\right\rangle \left\langle 24\right\rangle \left\langle 43\right\rangle \left\langle 31\right\rangle \right)\nonumber \\
 & -k_{s_{+}}\left(\left\langle 12\right\rangle \left\langle 23\right\rangle \left\langle 34\right\rangle \left\langle 41\right\rangle \right)^{2}-k_{s_{-}}\left(\left\langle 12\right\rangle \left\langle 24\right\rangle \left\langle 43\right\rangle \left\langle 31\right\rangle \right)^{2}\nonumber \\
 & -16\left(k_{a_{+}}\left\langle 12\right\rangle \left\langle 23\right\rangle \left\langle 34\right\rangle \left\langle 41\right\rangle +k_{a_{-}}\left\langle 12\right\rangle \left\langle 24\right\rangle \left\langle 43\right\rangle \left\langle 31\right\rangle \right)\left\langle 12\right\rangle \left\langle 23\right\rangle \left\langle 24\right\rangle \left\langle 34\right\rangle \left[21\right\rangle ]~,
\end{alignat}
}

\noindent \textcolor{black}{where
\begin{alignat}{1}
k_{c}= & (\left(k_{1}+k_{2}+k_{I}\right)\left(k_{3}+k_{4}+k_{I}\right)\left(k_{1}-k_{2}+k_{I}\right)\left(k_{3}-k_{4}+k_{I}\right)\nonumber \\
 & +\left(k_{1}+k_{2}-k_{I}\right)\left(k_{3}+k_{4}-k_{I}\right)\left(k_{1}-k_{2}-k_{I}\right)\left(k_{3}-k_{4}-k_{I}\right))\nonumber \\
 & (\left(k_{1}+k_{2}+k_{I}\right)\left(k_{3}+k_{4}+k_{I}\right)\left(k_{1}-k_{2}+k_{I}\right)\left(-k_{3}+k_{4}+k_{I}\right)\nonumber \\
 & +\left(k_{1}+k_{2}-k_{I}\right)\left(k_{3}+k_{4}-k_{I}\right)\left(k_{1}-k_{2}-k_{I}\right)\left(-k_{3}+k_{4}-k_{I}\right))\nonumber \\
+ & 8k_{I}^{4}\left(k_{1}+k_{2}+k_{3}+k_{4}\right)^{2}\left(k_{1}-k_{2}+k_{I}\right)\left(k_{1}-k_{2}-k_{I}\right)\;,
\end{alignat}
\begin{align}
k_{s_{\pm}}= & (\left(k_{1}+k_{2}+k_{I}\right)\left(k_{3}+k_{4}+k_{I}\right)\left(k_{1}-k_{2}+k_{I}\right)\left(\pm k_{3}\mp k_{4}+k_{I}\right)\nonumber \\
 & +\left(k_{1}+k_{2}-k_{I}\right)\left(k_{3}+k_{4}-k_{I}\right)\left(k_{1}-k_{2}-k_{I}\right)\left(\pm k_{3}\mp k_{4}-k_{I}\right))^{2}\;,
\end{align}
}

\textcolor{black}{
\begin{equation}
k_{a_{\pm}}=k_{I}^{4}\left(k_{1}+k_{2}+k_{3}+k_{4}\right)^{2}\left(\mp k_{1}\pm k_{2}-k_{3}+k_{4}\right)\;.
\end{equation}
}

\textcolor{black}{It turns out that all three types of terms are still
present, but it is clear that we cannot simply relate $\left\langle 1^{+}2^{+}3^{+}4^{+}\right\rangle _{s}$
to $\left\langle 1^{+}2^{+}3^{+}4^{+}\right\rangle _{s}^{YM}\left\langle 1^{+}2^{+}4^{+}3^{+}\right\rangle _{s}^{YM}$
by adding kinematic factor. Moreover, we cannot write $\left\langle 12\right\rangle \left\langle 23\right\rangle \left\langle 34\right\rangle \left\langle 41\right\rangle $
and $\left\langle 12\right\rangle \left\langle 24\right\rangle \left\langle 43\right\rangle \left\langle 31\right\rangle $
in terms of $\left\langle 1^{+}2^{+}3^{+}4^{+}\right\rangle _{s}^{YM}$
and $\left\langle 1^{+}2^{+}4^{+}3^{+}\right\rangle _{s}^{YM}$. Therefore
we have demonstrated that the KLT-like relations are even more non-trivial
then those in previous section when both sides are in inflation. This
proves the existence of the new features mentioned above. As a remark,
here we even do not have a clear picture on four-point interactions,
and whether adding up channels can cause extra simplifications is
more non-trivial than that in previous sections. Since the relation
becomes much less clear when both sides are in inflation, in the following
analysis we stick with the interpretation in terms of flat spacetime
amplitudes.}

\section{Behaviour of the Four-Point Functions\label{sec:4}}

\begin{figure}
\begin{centering}
\includegraphics[width=0.8\paperwidth]{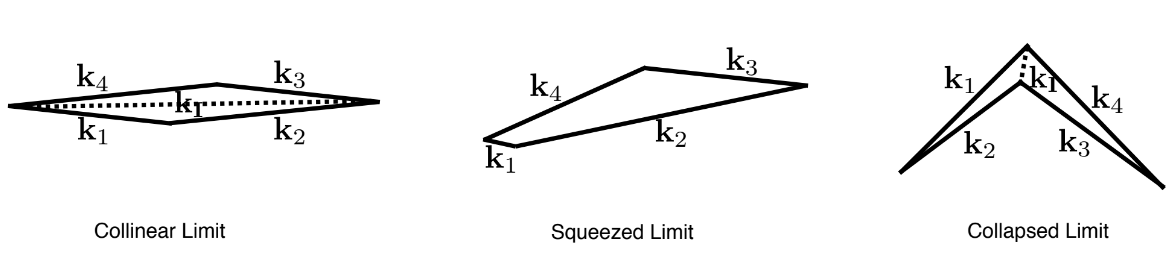} 
\par\end{centering}
\caption{Momentum configurations in different limits}

\label{fig:2} 
\end{figure}

We still need interpretations for the extra terms. To make better
sense of them, we study their behaviour in several limits, namely
the collinear limit, squeezed limit and collapsed limit, see Figure
\ref{fig:2}. In previous literature, many interesting properties
of (mostly scalar) correlation functions are found in such limits
\citep{Maldacena:2002vr,Creminelli:2004yq,Seery:2008ax,Li:2008gg,Assassi:2012zq,Senatore:2012wy,Berezhiani:2013ewa,Hinterbichler:2013dpa,Fu:2015vja,Chu:2018ovy}.
We expect similar properties can be found similarly. From this, we
can identify the correspondence between such properties and types
of terms. As mentioned above, we need to choose the anomalous term
which mostly facilitate our calculations. Here we try our best to
fix the choice $\left[21\right\rangle $ and only work out cases of
some independent combinations of helicities. The results can be easily
generalized to all combinations with other choices of anomalous terms.

\subsection{Collinear Limit}

Here we set the external momentum vectors to \textcolor{black}{align}
on the same straight line. Then the 3-momentum conservation $\mathbf{k}_{I}=-\mathbf{k}_{1}-\mathbf{k}_{2}=\mathbf{k}_{3}+\mathbf{k}_{4}$
implies $k_{1}+k_{2}+k_{I}\rightarrow0$ and $k_{3}+k_{4}-k_{I}\rightarrow0$
in the sense of analytic continuation. Here we maintain the choice
of direction of $k_{I}$ as above, and one can replace $k_{I}$ with
$-k_{I}$. The limits look like energy conservation, but keep in mind
that $k$ is just magnitudes of momentum and is not related to energy
in general. Therefore, we are not recovering the full result in flat
spacetime. \textcolor{black}{As a remark, we cannot really construct
the flat spacetime limit like the one in \cite{Raju:2012zs,Raju:2012zr}.
It is a price of that we greatly simplify the computation in previous
section by throwing away the information of energy of the internal
graviton. Even if we can construct the flat spacetime limit, here
we cannot demonstrate the well-known behaviour of full flat spacetime
amplitudes since we are not adding up channels in our previous results.}
\begin{itemize}
\item All $+$ 
\end{itemize}
From Equation (\ref{eq:24}), we can already see the correlation function
vanish. One interesting point is that if we only take the limit $k_{1}+k_{2}+k_{3}+k_{4}\rightarrow0$,
the correlation function is still not zero, but the anomalous term
disappears. 
\begin{itemize}
\item Three $+$ and One $-$ 
\end{itemize}
Under such limit we have $\left\langle 12\right\rangle \left[12\right]=\left(k_{1}+k_{2}+k_{I}\right)\left(k_{1}+k_{2}-k_{I}\right)=0$,
so $\left\langle 12\right\rangle =0$ or $\left[12\right]=0$. If
$\left\langle 12\right\rangle =0$, when we consider $\left\langle 1^{+}2^{-}3^{+}4^{+}\right\rangle _{s}$,
the anomalous term vanishes. It is also clear that $K_{c}$ and $K_{s_{\pm}}$
vanish after flipping the helicity. Therefore $\left\langle 1^{+}2^{-}3^{+}4^{+}\right\rangle _{s}=0$.
Changing the choice of anomalous term from $\left[21\right\rangle $
to $\left[12\right\rangle $ in Equation (\ref{eq:28}), we also have
$\left\langle 1^{-}2^{+}3^{+}4^{+}\right\rangle _{s}=0$. Similarly
if $\left[12\right]=0$, we have $\left\langle 1^{+}2^{-}3^{-}4^{-}\right\rangle _{s}=\left\langle 1^{-}2^{+}3^{-}4^{-}\right\rangle _{s}=0$.
Since such correlation functions become their complex conjugate when
all external helicities are flipped \citep{Fu:2015vja}, the conclusions
are true for all cases.

Here we get similar behaviour as that in flat spacetime. It is because
for amplitudes, if the internal graviton also goes on-shell, the conditions
of energy conservation at each vertex and the limits here become equivalent
to each other. Then it is expected that we recover some special cases
in flat spacetime, that is, vanishing amplitudes remaining vanishing. 
\begin{itemize}
\item Two $+$ and Two $-$ 
\end{itemize}
If we flip one external graviton from $+$ to $-$ at each vertex,
it is easy to see that only the square terms are dominant with the
same reason as above. For example, 
\begin{equation}
\left\langle 1^{+}2^{-}3^{-}4^{+}\right\rangle _{s}=\frac{\left(\left\langle 12\right]\left[23\right]\left[34\right\rangle \left\langle 41\right\rangle \right)^{2}}{k_{2}^{2}k_{3}^{2}}~.
\end{equation}

The result here seems to contradict to what we have in flat spacetime,
but it is just because we have not considered the contributions from
other channels. For example, if we flip helicities of two external
gravitons at one vertex i.e. $\left\langle 1^{-}2^{-}3^{+}4^{+}\right\rangle _{s}$
and $\left\langle 1^{+}2^{+}3^{-}4^{-}\right\rangle _{s}$, all three
types of terms are important. However, to consider the full contribution
we also need to consider the scalar parts. We will leave this for
a future work.

From above, we can see how the signs in the prefactors of spinor products
control the helicity structures. Such behaviour is not fully clear
in flat spacetime.

\subsection{Squeezed Limit}

Here we let one external graviton to be soft. For example, we consider
the limit $k_{1}\rightarrow0$. The condition of momentum conservation
becomes $\mathbf{k}_{2}+\mathbf{k}_{I}=0$ and thus $k_{2}=-k_{I}$.
The sign here depends on how we define the direction of $k_{I}$,
but the results are the same for all cases.

It is clear that for all combinations of helicities, the only dominant
term is the anomalous term if we keep the choice $\left[21\right\rangle $.
For instance, 
\begin{align}
\left\langle 1^{+}2^{+}3^{+}4^{+}\right\rangle _{s}= & \frac{\left(k_{2}+k_{3}+k_{4}\right)^{2}}{16\left(k_{1}k_{2}k_{3}k_{4}\right)^{2}}\nonumber \\
 & \left[\left(-k_{2}+k_{3}-k_{4}\right)\left\langle 12\right\rangle \left\langle 23\right\rangle \left\langle 34\right\rangle \left\langle 41\right\rangle +\left(k_{2}+k_{3}-k_{4}\right)\left\langle 12\right\rangle \left\langle 24\right\rangle \left\langle 43\right\rangle \left\langle 31\right\rangle \right]\left\langle 12\right\rangle \left\langle 23\right\rangle \left\langle 24\right\rangle \left\langle 34\right\rangle \left[21\right\rangle ~.
\end{align}

Note that $\left[21\right\rangle \left[12\right\rangle =\left(k_{I}+k_{1}-k_{2}\right)\left(k_{I}-k_{1}+k_{2}\right)=0$,
so either $\left[21\right\rangle $ or $\left[12\right\rangle $ vanishes.
We look at the non-vanishing $\left\langle 1^{+}2^{+}3^{+}4^{+}\right\rangle _{s}$
i.e. $\left[21\right\rangle \neq0$. Then by Equation (\ref{eq:35})
\begin{align}
\left\langle 1^{+}2^{+}3^{+}4^{+}\right\rangle _{s} & =\frac{\left(k_{2}+k_{3}+k_{4}\right)^{2}}{16\left(k_{1}k_{2}k_{3}k_{4}\right)^{2}}\left[\left(-k_{2}+k_{3}-k_{4}\right)\left\langle 12\right\rangle \left\langle 23\right\rangle \left\langle 34\right\rangle \left\langle 41\right\rangle +\left(k_{2}+k_{3}-k_{4}\right)\left\langle 12\right\rangle \left\langle 24\right\rangle \left\langle 43\right\rangle \left\langle 31\right\rangle \right]^{2}\nonumber \\
 & =\frac{\left(k_{2}+k_{3}+k_{4}\right)^{2}}{16\left(k_{1}k_{2}k_{3}k_{4}\right)^{2}}\left(\left\langle 12\right\rangle \left\langle 23\right\rangle \left\langle 24\right\rangle \left\langle 34\right\rangle \left[21\right\rangle \right)^{2}~.
\end{align}

Here the anomalous term and other types of terms can be converted
into each other. Therefore the non-trivial relation between anomalous
terms and other terms can be recovered in this limit. As a check,
when we also take $k_{2}\rightarrow0$, which implies $k_{I}\rightarrow0$,
we really recover the grouping of terms in collapsed limit.

Since $\left|1\right\rangle $ has order $\sqrt{k_{1}}$, both the
numerator and the denominator have order $k_{1}^{2}$ and the expression
is indeed finite. Note that in this limit the relation $k_{1}+k_{2}+k_{I}\rightarrow0$
still holds, so either $\left\langle 12\right\rangle $ or $\left[12\right]$
vanishes. Therefore for $s$ channel, at most two of the configurations
$1^{+}2^{+},\:1^{-}2^{+},\:1^{+}2^{-},\:1^{-}2^{-}$ can lead to non-zero
contributions.

Interestingly, under this limit we can find a relation between the
four-point functions and the three-point functions, including scalar
parts. Under this limit, there are simple consistency relations for
scalar correlators \citep{Assassi:2012zq,Berezhiani:2013ewa,Creminelli:2004yq,Hinterbichler:2013dpa},
but these relations do not apply here since we are considering one
specific channel, instead of full contributions. Nevertheless, using
the expression in \citep{Seery:2008ax} we can still construct 
\begin{equation}
\frac{\left\langle \phi_{\mathbf{k}_{1}}\phi_{\mathbf{k}_{2}}\phi_{\mathbf{k}_{3}}\phi_{\mathbf{k}_{4}}\right\rangle _{s}'}{\left\langle \phi_{\mathbf{k}_{1}}\phi_{-\mathbf{k}_{1}}\right\rangle '}=-\frac{\partial}{\partial\left(k_{2}^{2}\right)}\left\langle \phi_{\mathbf{k}_{2}}\phi_{\mathbf{k}_{3}}\phi_{\mathbf{k}_{4}}\right\rangle '~,
\end{equation}

\noindent up to a numerical factor which is not important. Here $\left\langle \phi...\phi\right\rangle $
denote the scalar parts of corresponding correlation functions. Note
that here it is natural to choose the direction of $k_{I}$ to be
incoming to $34I$ vertex i.e. $k_{2}=k_{I}$ in order to construct
the three-point function. Now add back the tensor parts and we get
\begin{align}
\frac{\left\langle h_{\mathbf{k}_{1}}^{+}h_{\mathbf{k}_{2}}^{+}h_{\mathbf{k}_{3}}^{+}h_{\mathbf{k}_{4}}^{+}\right\rangle _{s}'}{\left\langle h_{\mathbf{k}_{1}}^{+}h_{\mathbf{-k}_{1}}^{+}\right\rangle '} & =-\frac{\left(k_{2}+k_{3}+k_{4}\right)^{2}}{16\left(k_{1}k_{2}k_{3}k_{4}\right)^{2}}\left(\left\langle 12\right\rangle \left\langle 23\right\rangle \left\langle 24\right\rangle \left\langle 34\right\rangle \left[21\right\rangle \right)^{2}\frac{\partial}{\partial\left(k_{2}^{2}\right)}\frac{16k_{2}^{2}k_{3}^{2}k_{4}^{2}\left\langle h_{\mathbf{k}_{2}}^{+}h_{\mathbf{k}_{3}}^{+}h_{\mathbf{k}_{4}}^{+}\right\rangle '}{\left\langle 23\right\rangle ^{2}\left\langle 34\right\rangle ^{2}\left\langle 42\right\rangle ^{2}\left(k_{2}+k_{3}+k_{4}\right)^{2}}\nonumber \\
 & =-\frac{\left[\left\langle 12\right\rangle \left[21\right\rangle \right]^{2}}{k_{1}^{2}}\frac{\partial}{\partial\left(k_{2}^{2}\right)}\left\langle h_{\mathbf{k}_{2}}^{+}h_{\mathbf{k}_{3}}^{+}h_{\mathbf{k}_{4}}^{+}\right\rangle '+K\left\langle h_{\mathbf{k}_{2}}^{+}h_{\mathbf{k}_{3}}^{+}h_{\mathbf{k}_{4}}^{+}\right\rangle '~,
\end{align}

\noindent where $K$ is a kinematic factor. Here it is interesting
to observe the cancellation of kinematic factors in the derivative
term. Note that $\frac{\left[\left\langle 12\right\rangle \left[21\right\rangle \right]^{2}}{k_{1}^{2}}\sim\epsilon_{ij}^{1}k_{i}^{2}k_{j}^{2}$,
so this relation has the same form as the consistency relation between
three-point and two-point functions \citep{Creminelli:2004yq,Maldacena:2002vr,Hinterbichler:2013dpa},
but with an extra term proportional to the three-point function due
to the presence of tensor parts in the three-point function.

\subsection{Collapsed Limit}

We take the collapsed limit $k_{I}\rightarrow0$ and consider leading
order terms in Equation (\ref{eq:28}). This also implies that $\left|k_{1}\right|\approx\left|k_{2}\right|$
and $\left|k_{3}\right|\approx\left|k_{4}\right|$ since the limit
forces $\mathbf{k}_{1}+\mathbf{k}_{2}=\mathbf{k}_{3}+\mathbf{k}_{4}\approx0$.
However, since the magnitudes are just approximately equal, we keep
them to be distinct. In this way we actually keep some higher order
terms implicitly.

It is clear that for all combinations of helicities the anomalous
terms become negligible while other terms remain important. However,
the remaining terms can be grouped into a simple expression by Schouten's
identity. For example, 
\begin{equation}
\left\langle 1^{+}2^{+}3^{+}4^{+}\right\rangle _{s}=\frac{\left(\left(k_{1}^{2}-k_{2}^{2}\right)\left(k_{3}^{2}-k_{4}^{2}\right)\right)^{2}}{128\left(k_{1}k_{2}k_{3}k_{4}k_{I}^{2}\right)^{2}}\left\langle 12\right\rangle ^{4}\left\langle 34\right\rangle ^{4}~.\label{eq:33}
\end{equation}

We then have a nice factorization. This can be understood as when
the helicity of internal graviton changes, we have $k_{I}\rightarrow-k_{I}$
but the expression is insensitive to this in the collapsed limit.
Therefore we can recover the multiplication of two three-point vertices.
It also shows that there exist relations between different types of
terms, while such relations are non-trivial in flat spacetime.

\begin{figure}
\begin{centering}
\includegraphics[width=0.8\paperwidth]{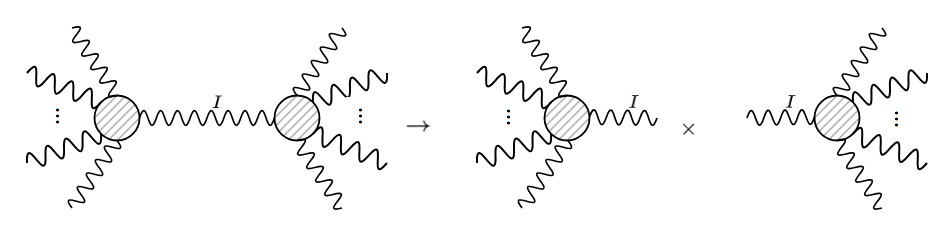} 
\par\end{centering}
\caption{Factorization of correlation function in collapsed limit into two
correlation functions in squeezed limit}

\label{fig:3} 
\end{figure}

It is worth mentioning that although we do not know the explicit forms
of higher-point functions, with the same logic we can factorize a
correlation function into product of two lower-point correlation functions
when one of internal graviton becomes soft \citep{Assassi:2012zq,Senatore:2012wy},
see Figure \ref{fig:3}. The soft internal gravitons in the two lower
point correlation functions have opposite 3-momenta and same helicity,
which is equivalent to the analytic continuation as we do for four-point
functions when we only consider one internal graviton. It is because
in such case both are equivalent to transforming $\left|k\right]\rightarrow\left|k\right\rangle $
and vice versa \citep{Maldacena:2011nz}. However, the result is the
same no matter what helicity and direction of momentum of the soft
internal graviton we choose in one of the correlation functions. Conventionally,
after the factorization, we can apply consistency relations in squeezed
limit as mentioned in last section to further simplify the correlation
functions. An example is given in Appendix \ref{sec:C}.

As a check, we let all momenta lie on the same plane and take $\left|k_{1}\right|=\left|k_{2}\right|$
and $\left|k_{3}\right|=\left|k_{4}\right|$. Then both numerator
and denominator seem to vanish. However, before taking so, we can
first approximate $k_{I}\approx\pm\left(\left|k_{1}\right|-\left|k_{2}\right|\right)\approx\pm\left(\left|k_{3}\right|-\left|k_{4}\right|\right)$
and $\left|2\right\rangle \rightarrow-\left|1\right],\left|4\right\rangle \rightarrow-\left|3\right]$
\citep{Maldacena:2011nz}. The signs depend on how we define the direction
of $k_{I}$, but the result is independent of the signs. Finally Equation
(\ref{eq:33}) becomes simply 
\begin{equation}
\left\langle 1^{+}2^{+}3^{+}4^{+}\right\rangle _{s}=32k_{1}^{2}k_{3}^{2}~.
\end{equation}

Note that it is non-zero only when the external helicities are the
same at each vertex. In other words, only $\left\langle 1^{+}2^{+}3^{+}4^{+}\right\rangle _{s},\left\langle 1^{-}2^{-}3^{+}4^{+}\right\rangle _{s},\left\langle 1^{+}2^{+}3^{-}4^{-}\right\rangle _{s}\left\langle 1^{-}2^{-}3^{-}4^{-}\right\rangle _{s}$
do not vanish. Therefore our calculation is consistent with the cases
with external scalars \citep{Seery:2008ax} and linear polarization
\citep{Fu:2015vja}. In general, if not all momenta lie on the same
plane, there is also an angular dependence from the configurations
of momenta, see Appendix \ref{sec:C}, but here it is conveniently
encoded into the spinor products.

The first order contribution in $k_{I}$ vanishes as mentioned before.
When we go to second order, the expression has become complicated
and contained anomalous terms.

To summarize, different types of terms have their peculiar roles and
properties in the whole correlation functions, while there exist non-trivial
relations between different types of terms which can be recovered
in certain limits. Such division of roles interestingly controls the
helicity structure and represents the distinction between de Sitter
spacetime and flat spacetime. Various properties of the correlation
functions also become transparent under our formalism.

\section{Conclusion\label{sec:5}}

In this paper, we generalize the inflationary spinor helicity formalism
in \citep{Maldacena:2011nz} to four-point functions. Through this
we derive a KLT-like relation, which contains some extra terms, including
terms that do not look like square of amplitudes, when compared to
what KLT relations in flat spacetime predict. \textcolor{black}{These
terms may be new features in de Sitter spacetime or inflation, which
make the KLT structure very unclear. Therefore, what we present here
is only some preliminary work on constructing a complete KLT-like
relation, and much further research on this direction can be developed.}
Interesting topics along this direction include 
\begin{itemize}
\item It remains interesting to seek for more interpretations and physical
meaning of those extra terms. There may be non-trivial relations between
different types of terms. In addition, it is important to find further
ways to simplify the anomalous terms, since so far we do not have
a rigorous proof to show that whether the anomalous terms are really
``anomalous'', or just a non-trivial form of square of amplitudes.
We have such possibility since our calculations can be facilitated
with certain choices of anomalous terms, showing that there are still
some other relations hidden between anomalous terms and other terms. 
\item It may be useful to consider the diagrams of other channels or permutations
of external gravitons. As the case in flat spacetime, the correlators
may be further simplified when we add up different contributions.
However, to do so we also need the information of the scalar parts
since they are different for different channels in general \citep{Seery:2008ax}.
Here we only focus on the tensor parts, but it would be nice to see
if such simplifications happen and new properties may be discovered
in this way. 
\item It is natural to investigate higher-than-four-point correlations.
It is expected to get the KLT relations for higher-point amplitudes
as parts of the relations. However, the interpretations of those extra
terms may change. In addition, it would be interesting if new types
of extra terms, especially anomalous terms appear in higher-point
functions. 
\item Here we only work with the simplest minimally coupled inflation model
with Einstein gravity. There is a large variety of inflation models
and modified gravity. One may investigate the applications of the
formalism here to other models. This generalization is highly non-trivial
since the factorization into scalar parts and tensor parts can only
applied to some specific models. 
\item So far we only focus on spinor helicity formalism, which may not be
a good method because the symmetries required by the formalism are
not all present in inflation. However, there are other symmetries
in de Sitter spacetime while they are not present in flat spacetime,
such as the conformal symmetries. One may use the conformal invariance
of correlators to enforce the forms of them, bypassing direct computations.
For example, this can help us tackle the tedious algebra of special
functions when we consider massive fields \citep{Arkani-Hamed:2015bza}.
This kind of work has been done for three-point functions from both
interactions purely between gravitons and between scalar and gravity
\citep{Maldacena:2011nz,Mata:2012bx}. It is interesting to know if
it works also for higher-point functions. In fact, similar work has
been done in, for example, \citep{Raju:2012zs} in the context of
AdS/CFT correspondence \citep{Maldacena:1997re,Maldacena:2002vr}. 
\end{itemize}
We hope to address some of the above issues in our further studies. 
\begin{acknowledgments}
We thank Henry Tye for many helpful discussions. This work is supported
in part by ECS Grant 26300316 and GRF Grant 16301917 from the Research
Grants Council of Hong Kong. 
\end{acknowledgments}

\appendix

\section{Spinor Helicity Formalism in Flat Spacetime\label{sec:A}}

The spinor helicity formalism is one of the methods to simplify amplitudes
in flat spacetime. For simplicity we use the gluon scattering as examples,
since $\mathrm{Gravity}=\mathrm{Gauge}^{2}$ is well-known in flat
spacetime and one can easily generalize the results to graviton scattering.
Although the below results are already well-known, we still need to
point out some key results in order to make comparisons to the case
in inflation.

The motivation to do spinor helicity formalism is that both gluon
and graviton amplitudes are too complicated if we only apply Feynman
rules \citep{Elvang:2013cua,PhysRevD.34.1749}. They involve thousands,
or even millions of products between polarization tensors and momenta,
even if we only consider low-point functions. However, spinor helicity
formalism reveals the hidden symmetry in these terms and helps us
group them into simple expressions. The complicated dependence on
momenta can also be encoded into simple spinor products. The formalism
is based on the simple fact \citep{Schwartz:2013pla} 
\begin{equation}
SO\left(1,3\right)\cong SU\left(2\right)\otimes SU\left(2\right)~.
\end{equation}

Therefore Lorentz vectors, tensors etc. can be decomposed into multiple
spinors. This seems not to be a simplification, but amazing it turns
out to have great power if we consider massless particles. We now
thus only consider lightlike momentum vectors. Here we mostly use
the symbols in \citep{Elvang:2013cua}.

\subsection{Spinors}

First consider the massless Dirac equation 
\begin{equation}
\cancel{p}u\left(p\right)=0~.
\end{equation}

The two independent solutions are denoted as 
\begin{equation}
u_{-}\left(p\right)=\left(\begin{array}{c}
\left|p\right]_{a}\\
0
\end{array}\right),\:u_{+}\left(p\right)=\left(\begin{array}{c}
0\\
\left|p\right\rangle _{\dot{a}}
\end{array}\right)~.
\end{equation}

The dotted and undotted indices are just the same as tensor notations.
We use Levi-Civita symbols to raise and lower indices 
\begin{equation}
\varepsilon^{ab}=\varepsilon^{\dot{a}\dot{b}}=-\varepsilon_{ab}=-\varepsilon_{\dot{a}\dot{b}}=\left(\begin{array}{cc}
0 & 1\\
-1 & 0
\end{array}\right)~.
\end{equation}

Then we further define 
\begin{equation}
\left[p\right|^{a}=\varepsilon^{ab}\left|p\right]_{b},\:\left\langle p\right|^{\dot{a}}=\varepsilon^{\dot{a}\dot{b}}\left|p\right\rangle _{\dot{b}}~,
\end{equation}
\begin{equation}
\left[pq\right]=\left[p\right|^{a}\left|q\right]_{a},\:\left\langle pq\right\rangle =\left\langle p\right|^{\dot{a}}\left|q\right\rangle _{\dot{a}}~.
\end{equation}

To facilitate the calculations, sometimes one may use the explicit
forms of the spinors. For example, 
\begin{equation}
\left|p\right]_{a}=\left(\begin{array}{c}
\frac{-p_{1}+ip_{2}}{\sqrt{p_{0}+p_{3}}}\\
\sqrt{p_{0}+p_{3}}
\end{array}\right),\left|p\right\rangle _{\dot{a}}=\left(\begin{array}{c}
\sqrt{p_{0}+p_{3}}\\
\frac{p_{1}+ip_{2}}{\sqrt{p_{0}+p_{3}}}
\end{array}\right)~,
\end{equation}

\noindent where $p_{0}$ to $p_{3}$ are the components of momentum
4-vector. Usually we only need to work on angle bracket and square
bracket products, which do not contain spinor indices. Therefore we
ignore the spinor indices, while one can add back them straightforwardly.
Using the definitions, one can derive the following 
\begin{equation}
\left\langle pq\right\rangle =-\left\langle qp\right\rangle ,\:\left[pq\right]=-\left[qp\right],\:\left\langle pp\right\rangle =\left[pp\right]=0~,
\end{equation}
\begin{equation}
\left\langle pq\right\rangle \left[pq\right]=-2p\cdot q=2pq-2\mathbf{p}\cdot\mathbf{q}~,
\end{equation}
\begin{equation}
\left\langle p\right|P\left|q\right]=P_{\mu}\left\langle p\right|\gamma^{\mu}\left|q\right]=-\left\langle pP\right\rangle \left[Pq\right]~,
\end{equation}
\begin{equation}
\left\langle 1\right|\gamma^{\mu}\left|2\right]\left\langle 3\right|\gamma_{\mu}\left|4\right]=2\left\langle 13\right\rangle \left[24\right]~,\label{eq:44}
\end{equation}
\begin{equation}
\left|i\right\rangle \left\langle jk\right\rangle +\left|j\right\rangle \left\langle ki\right\rangle +\left|k\right\rangle \left\langle ij\right\rangle =0~,\label{eq:45}
\end{equation}

\begin{equation}
\left|-p\right]=i\left|p\right],\:\left|-p\right\rangle =i\left|p\right\rangle ~.
\end{equation}

Here $\gamma^{\mu}$ are the conventional gamma matrices and we use
the shorthand notation of $\left|p_{n}\right\rangle =\left|n\right\rangle $.
The products with gamma matrices are defined as 
\begin{equation}
\left\langle p\right|\gamma^{\mu}\left|k\right]=\left(\begin{array}{cc}
0 & \left\langle p\right|\end{array}\right)\left(\begin{array}{cc}
0 & \sigma^{\mu}\\
\bar{\sigma}^{\mu} & 0
\end{array}\right)\left(\begin{array}{c}
\left|k\right]\\
0
\end{array}\right)~.
\end{equation}

Equation (\ref{eq:44}) is then the well-known Fierz's Identity. Equation
(\ref{eq:45}) is called the Schouten's Identity.

A very useful trick can be done by 4-momentum conservation. If $p_{1}^{\mu}+p_{2}^{\mu}+...+p_{n}^{\mu}=0$,
we have 
\begin{equation}
\stackrel[i=1]{n}{\sum}\left\langle ji\right\rangle \left[ik\right]=-\left\langle j\right|1+2+...+n\left|k\right]=0~.\label{eq:57}
\end{equation}

For $n=3$, using the trick we can also derive the 3-particle special
kinematics: 
\begin{equation}
\mathrm{Either}\:\left\langle 12\right\rangle ,\,\left\langle 23\right\rangle ,\,\left\langle 31\right\rangle \:\mathrm{or}\:\left[12\right],\,\left[23\right],\,\left[31\right]=0~.\label{eq:58}
\end{equation}

Note that Equations (\ref{eq:57}) and (\ref{eq:58}) are no longer
true in inflation.

\subsection{Vectors}

Now we can decompose vectors into bispinors. First observe that for
a vector $p^{\mu}$, $\bar{\sigma}_{\mu}p^{\mu}$ also transform as
a vector. To decompose it, one can check that 
\begin{equation}
\bar{\sigma}_{\mu}p^{\mu}=-\left|p\right\rangle \left[p\right|~.
\end{equation}

For gluon amplitudes, we usually consider products among polarization
vectors and momenta only. We use circular polarization of gluons and
impose the gauge conditions $\epsilon_{s\mu}^{*}\epsilon_{s'}^{\mu}=-\delta_{ss'}$
and $p_{\mu}\epsilon^{\mu}=0$. Then a more intuitive form of polarization
vectors can be used: 
\begin{equation}
\epsilon_{+}^{\mu}\left(p\right)=-\frac{\left\langle p\right|\gamma^{\mu}\left|q\right]}{\sqrt{2}\left[qp\right]},\:\epsilon_{-}^{\mu}\left(p\right)=-\frac{\left\langle q\right|\gamma^{\mu}\left|p\right]}{\sqrt{2}\left\langle qp\right\rangle }~,\label{eq:50}
\end{equation}

\noindent where $q\neq p$ is called the reference spinor. Due to
redundancy in Yang-Mills theory, the $q$ is arbitrary except $q\neq p$
but the amplitudes are always independent of $q$. One can easily
check that Equation (\ref{eq:50}) satisfies all gauge conditions
above for all $q\neq p$. Here comes the greatest power of spinor
helicity formalism. We can always choose the $q$ that makes our calculations
convenient. When we make the smart choice, dramatic simplifications
will happen and they reveal the symmetry among the huge amounts of
terms in amplitudes.

\subsection{MHV Amplitudes}

With the above tools, we can rewrite various amplitudes with definite
helicities into spinor product form. Below we denote the colour-stripped
amplitudes to be $A\left(\mathrm{helicities}\right)$. Keep in mind
that we only focus on the part with products among polarization vectors
and momenta.

Let us consider the case when all $n$ gluons have $+$ polarizations.
There are only three-point vertices and four-point vertices in Yang-Mills
theory. From their Feynman rules \citep{Schwartz:2013pla}, we see
that there are at most $n-2$ momenta since each vertex is associated
with at most one momentum vector, while there are $n$ polarization
vectors. Then there must be two polarization vectors contracting with
each other. Now by setting reference spinors of all polarization vectors
being the same, it is clear that products between two polarization
vectors become zero. Therefore, 
\begin{equation}
A\left(++...++\right)=0~.
\end{equation}

This beautiful result seems to be simple in spinor helicity formalism,
but it is highly non-trivial if we only consider the terms from Feynman
rules. With similar arguments, one can also prove that 
\begin{equation}
A\left(--...--\right)=0~,
\end{equation}

\noindent and 
\begin{equation}
A\left(-+...++\right)=A\left(+-...--\right)=0~,
\end{equation}
for more than three gluons. Therefore the first non-zero amplitudes
with most gluons having same helicities are the amplitudes with 2
gluons having opposite helicities to other gluons, which is called
maximally helicity violating (MHV) amplitudes. Using spinor helicity
formalism and other modern tools such as BCFW recursion relations,
one can easily find an amazingly simple form of the amplitudes known
as Parke-Taylor formula \citep{Elvang:2013cua,Schwartz:2013pla} 
\begin{equation}
A\left(1^{-}...i^{+}...j^{+}...n^{-}\right)=\frac{\left\langle ij\right\rangle ^{4}}{\left\langle 12\right\rangle \left\langle 23\right\rangle ...\left\langle n1\right\rangle }~,
\end{equation}

\noindent where the external legs of $1$ to $n$ are arranged clockwisely
in the Feynman diagrams. Its compactness is extremely attractive since
for very high-point amplitudes, it reduces the enormous amounts of
terms that even computers cannot handle \citep{Zee:2003mt} into an
easily computable expression. Therefore we have strong hope of finding
similar relations in inflation. However, this kind of work turns out
to be difficult due to symmetry breaking, and indeed the conclusions
in this subsection are not true in inflation.

\section{Three-Point Graviton Correlators\label{sec:B}}

We demonstrate how to simplify \textcolor{black}{the tensor parts
of} graviton three-point functions by the above formalism in details.
Keep in mind that we need to specify the helicity of each particle
in our formalism. For three-point functions, there is only one diagram.

First we calculate $\left\langle 1^{+}2^{+}3^{+}\right\rangle $.
Just substituting and playing around with the above formulas, we find
\begin{equation}
k_{i}^{2}k_{j}^{2}\epsilon_{ij}^{1}\epsilon_{kl}^{2}\epsilon_{kl}^{3}=\frac{\left\langle 12\right\rangle ^{2}\left\langle 23\right\rangle ^{2}\left\langle 31\right\rangle ^{2}}{16k_{1}^{2}k_{2}^{2}k_{3}^{2}}\left(k_{2}+k_{3}-k_{1}\right)^{2}~,
\end{equation}
\begin{equation}
\epsilon_{ij}^{1}k_{l}^{3}\epsilon_{li}^{2}k_{m}^{2}\epsilon_{mj}^{3}=-\frac{\left\langle 12\right\rangle ^{2}\left\langle 23\right\rangle ^{2}\left\langle 31\right\rangle ^{2}}{16k_{1}^{2}k_{2}^{2}k_{3}^{2}}\left(k_{1}+k_{2}-k_{3}\right)\left(k_{1}+k_{3}-k_{2}\right)~.
\end{equation}

Since the correlator does not change when we permute $1,2,3$, performing
the cyclic sum directly we get 
\begin{equation}
\left\langle 1^{+}2^{+}3^{+}\right\rangle =\frac{\left\langle 12\right\rangle ^{2}\left\langle 23\right\rangle ^{2}\left\langle 31\right\rangle ^{2}}{16k_{1}^{2}k_{2}^{2}k_{3}^{2}}\left(k_{1}+k_{2}+k_{3}\right)^{2}~.
\end{equation}

Here many interesting things already happen. First, for each term
in the cyclic sum, we have expressions as complex as the final expression.
This shows that some symmetries are hidden in those terms and our
formalism really simplifies something. Secondly, although $A\left(+++\right)=0$
in flat spacetime, if we assume it is non-zero by little group scaling
\citep{Elvang:2013cua,Schwartz:2013pla} it has the form $\left\langle 12\right\rangle \left\langle 23\right\rangle \left\langle 31\right\rangle $.
Therefore the pattern of $\mathrm{Gravity}=\mathrm{Gauge}^{2}$ for
three-point functions remains the same in inflation.

The last thing, which is the most significant, is that $\left\langle 1^{+}2^{+}3^{+}\right\rangle \neq0$
in general, but only vanishes when the 3 gravitons are on-shell and
energy is conserved i.e. $k_{1}+k_{2}+k_{3}=0$. It is largely different
from the case in flat spacetime. From this, we would also expect all
the correlators $\left\langle ++...++\right\rangle ,\left\langle -+...++\right\rangle ,\left\langle +-...--\right\rangle ,\left\langle --...--\right\rangle $
to become non-zero. Since many beautiful results in flat spacetime,
such as the Parke-Taylor formula, are largely based on those vanishing
amplitudes, it is clear that the results become much more complicated
in inflation. This completely reveals the difficulties of doing calculations
in inflation due to symmetry breaking. It also implies that we really
cannot choose the reference spinors we like, otherwise we would get
back those vanishing amplitudes or correlators, which lead to contradictions.

We can still have a look to other combinations of helicities. For
instance, we consider $\left\langle 1^{+}2^{+}3^{-}\right\rangle $.
Since the correlator changes when we permute $1,2,3$, we must calculate
the six terms separately. 
\begin{equation}
k_{i}^{2}k_{j}^{2}\epsilon_{ij}^{1}\epsilon_{kl}^{2}\epsilon_{kl}^{3}=\frac{\left\langle 12\right\rangle ^{2}\left\langle 23\right]^{2}\left[31\right\rangle ^{2}}{16k_{1}^{2}k_{2}^{2}k_{3}^{2}}\left(k_{1}+k_{3}-k_{2}\right)^{2}~,
\end{equation}
\begin{equation}
k_{i}^{3}k_{j}^{3}\epsilon_{ij}^{2}\epsilon_{kl}^{3}\epsilon_{kl}^{1}=\frac{\left\langle 12\right\rangle ^{2}\left\langle 23\right]^{2}\left[31\right\rangle ^{2}}{16k_{1}^{2}k_{2}^{2}k_{3}^{2}}\left(k_{2}+k_{3}-k_{1}\right)^{2}~,
\end{equation}
\begin{equation}
k_{i}^{1}k_{j}^{1}\epsilon_{ij}^{3}\epsilon_{kl}^{1}\epsilon_{kl}^{2}=\frac{\left\langle 12\right\rangle ^{2}\left\langle 23\right]^{2}\left[31\right\rangle ^{2}}{16k_{1}^{2}k_{2}^{2}k_{3}^{2}}\left(k_{1}+k_{2}+k_{3}\right)^{2}~,
\end{equation}
\begin{equation}
\epsilon_{ij}^{1}k_{l}^{3}\epsilon_{li}^{2}k_{m}^{2}\epsilon_{mj}^{3}=\frac{\left\langle 12\right\rangle ^{2}\left\langle 23\right]^{2}\left[31\right\rangle ^{2}}{16k_{1}^{2}k_{2}^{2}k_{3}^{2}}\left(k_{1}+k_{2}+k_{3}\right)\left(k_{2}+k_{3}-k_{1}\right)~,
\end{equation}
\begin{equation}
\epsilon_{ij}^{2}k_{l}^{1}\epsilon_{li}^{3}k_{m}^{3}\epsilon_{mj}^{1}=\frac{\left\langle 12\right\rangle ^{2}\left\langle 23\right]^{2}\left[31\right\rangle ^{2}}{16k_{1}^{2}k_{2}^{2}k_{3}^{2}}\left(k_{1}+k_{2}+k_{3}\right)\left(k_{1}+k_{3}-k_{2}\right)~,
\end{equation}
\begin{equation}
\epsilon_{ij}^{3}k_{l}^{2}\epsilon_{li}^{1}k_{m}^{1}\epsilon_{mj}^{2}=-\frac{\left\langle 12\right\rangle ^{2}\left\langle 23\right]^{2}\left[31\right\rangle ^{2}}{16k_{1}^{2}k_{2}^{2}k_{3}^{2}}\left(k_{2}-k_{1}-k_{3}\right)\left(k_{1}-k_{2}-k_{3}\right)~.
\end{equation}

Again they look quite complicated, but finally we get 
\begin{equation}
\left\langle 1^{+}2^{+}3^{-}\right\rangle =\frac{\left\langle 12\right\rangle ^{2}\left\langle 23\right]^{2}\left[31\right\rangle ^{2}}{16k_{1}^{2}k_{2}^{2}k_{3}^{2}}\left(k_{1}+k_{2}-k_{3}\right)^{2}~.
\end{equation}

As above the final sum is more compact. Here we see one advantage
of crossing between angle brackets and square brackets. If the helicity
of a graviton changes from $+$ to $-$, we just change the corresponding
part of the correlator by $\left.\right\rangle \rightarrow\left.\right]$
and $k\rightarrow-k$ and vice versa. Therefore after we calculate
the correlator for one combination of helicities, it can be easily
generalized to other combinations. Since higher-point correlators
are formed by three-point vertices, the transformation is true in
general.

Although it has different form from that in flat spacetime due to
presence of crossing products, one can show that $\mathrm{Gravity}=\mathrm{Gauge}^{2}$
still works. Note that we are just comparing the above to gluon amplitudes
in flat spacetime, since the corresponding Yang-Mills theory is not
naturally present in our simple inflation model.

\section{Collapsed Limit of Four-Point Functions from Conventional Method\label{sec:C}}

Here we explicitly compute the expression of the four-point \textcolor{black}{function}
in Section \ref{sec:4} in \textcolor{black}{collapsed} limit, using
conventional method. We set up a coordinate system and use the explicit
expressions of polarization tensors. To show how our above results
are consistent with previous literature, we also compute the scalar
part. First we set up a spherical coordinate system: 
\[
\mathbf{k}_{I}=k_{I}\left(0,0,1\right),\;\mathbf{k}_{1}=k_{1}\left(\sin\theta_{1},0,\cos\theta_{1}\right),\;\mathbf{k}_{2}=k_{2}\left(-\sin\phi_{1},0,-\cos\phi_{1}\right)~,
\]
\begin{equation}
\mathbf{k}_{3}=k_{3}\left(\cos\alpha\sin\theta_{2},\sin\alpha\sin\theta_{2},\cos\theta_{2}\right),\;\mathbf{k}_{4}=k_{4}\left(-\cos\alpha\sin\phi_{2},-\sin\alpha\sin\phi_{2},-\cos\phi_{2}\right)~.
\end{equation}

Therefore when $\alpha=0,\pi$, all momenta lie on the same plane.
From Equation (\ref{eq16}), for momentum pointing to $z$ direction
such as $\mathbf{k}_{I}$, we have 
\begin{equation}
\epsilon_{ij}^{\pm}=\left(\begin{array}{ccc}
1 & \pm i & 0\\
\pm i & -1 & 0\\
0 & 0 & 0
\end{array}\right)~,
\end{equation}

\noindent and the polarization tensors for other momenta can be easily
obtained by applying \textcolor{black}{corresponding} rotation matrices
to above. Now we take the \textcolor{black}{collapsed} limit $k_{I}\rightarrow0$.
Applying the method in \citep{Assassi:2012zq,Senatore:2012wy}, we
first write down 
\begin{equation}
\left\langle h_{\mathbf{k}_{1}}^{s_{1}}h_{\mathbf{k}_{2}}^{s_{2}}h_{\mathbf{k}_{3}}^{s_{3}}h_{\mathbf{k}_{4}}^{s_{4}}\right\rangle _{s}'=\underset{s}{\sum}\left\langle h_{\mathbf{k}_{I}}^{s}h_{-\mathbf{k}_{I}}^{s}\right\rangle '\frac{\left\langle h_{\mathbf{k}_{I}}^{s}h_{\mathbf{k}_{1}}^{s_{1}}h_{\mathbf{k}_{2}}^{s_{2}}\right\rangle '\left\langle h_{-\mathbf{k}_{I}}^{s}h_{\mathbf{k}_{3}}^{s_{3}}h_{\mathbf{k}_{4}}^{s_{4}}\right\rangle '}{\left\langle h_{\mathbf{k}_{I}}^{s}h_{\mathbf{-k}_{I}}^{s}\right\rangle '\left\langle h_{\mathbf{k}_{I}}^{s}h_{-\mathbf{k}_{I}}^{s}\right\rangle '}~.
\end{equation}

Using the consistency relation for three-point functions in squeezed
limit \citep{Maldacena:2002vr,Creminelli:2004yq,Hinterbichler:2013dpa}:
\[
\frac{\left\langle h_{\mathbf{k}_{I}}^{s}h_{\mathbf{k}_{1}}^{s_{1}}h_{\mathbf{k}_{2}}^{s_{2}}\right\rangle '}{\left\langle h_{\mathbf{k}_{I}}^{s}h_{\mathbf{-k}_{I}}^{s}\right\rangle '}=-\epsilon_{ij}^{s}k_{i}^{1}k_{j}^{2}\frac{\partial}{\partial\left(k_{1}^{2}\right)}\left\langle h_{\mathbf{k}_{1}}^{s_{1}}h_{\mathbf{k}_{2}}^{s_{2}}\right\rangle '~,
\]

\noindent we arrive 
\begin{align}
\left\langle h_{\mathbf{k}_{1}}^{s_{1}}h_{\mathbf{k}_{2}}^{s_{2}}h_{\mathbf{k}_{3}}^{s_{3}}h_{\mathbf{k}_{4}}^{s_{4}}\right\rangle _{s}' & =\underset{s}{\sum}\left\langle h_{\mathbf{k}_{I}}^{s}h_{-\mathbf{k}_{I}}^{s}\right\rangle '\frac{\partial}{\partial\left(k_{1}^{2}\right)}\left\langle h_{\mathbf{k}_{1}}^{s_{1}}h_{\mathbf{k}_{2}}^{s_{2}}\right\rangle '\frac{\partial}{\partial\left(k_{3}^{2}\right)}\left\langle h_{\mathbf{k}_{3}}^{s_{3}}h_{\mathbf{k}_{4}}^{s_{4}}\right\rangle '\left(\epsilon_{ij}^{s}k_{i}^{1}k_{j}^{2}\epsilon_{lm}^{s}k_{l}^{3}k_{m}^{4}\right)\nonumber \\
 & =\left\langle h_{\mathbf{k}_{I}}^{\pm}h_{-\mathbf{k}_{I}}^{\pm}\right\rangle '\frac{\partial}{\partial\left(k_{1}^{2}\right)}\left\langle h_{\mathbf{k}_{1}}^{s_{1}}h_{\mathbf{k}_{2}}^{s_{2}}\right\rangle '\frac{\partial}{\partial\left(k_{3}^{2}\right)}\left\langle h_{\mathbf{k}_{3}}^{s_{3}}h_{\mathbf{k}_{4}}^{s_{4}}\right\rangle '\left(2k_{1}^{2}k_{3}^{2}\cos2\alpha\right)~,
\end{align}

\noindent where we have used the fact that in collasped limit, $\theta_{1},\theta_{2},\phi_{1},\phi_{2}\rightarrow\frac{\pi}{2}$
\citep{Seery:2008ax}. Recall that a two-point function does not have
tensor parts. It vanishes when the two helicities are opposite and
independent of whether the two helicities are $++$ or $--$. Therefore
it reduces to the result in Section \ref{sec:4} as expected, but
there is a difference in the numerical factor since some products
between polarization tensors become numerical constants by the normalization
condition.

 \bibliographystyle{utphys}
\bibliography{gwave_helicity}

\end{document}